# Post-radiation phenomena in thermally treated Kr matrices

E. Savchenko[1], I. Khyzhniy[1], S. Uyutnov[1], M. Bludov[1], A. Ponomaryov[2] and V. Bondybey[3]

[1]B. Verkin Institute for Low Temperature Physics & Engineering NASU, Kharkiv 61103, Ukraine,

[2]Helmholtz Zentrum Dresden-Rossendorf, 01328 Dresden, Germany

[3]Lehrstuhl für Physikalische Chemie II TUM, Garching b. Munich 85747, Germany

**Abstract**

The effect of thermal treatment on relaxation phenomena in Kr matrices irradiated with a low energy electron beam has been studied. The experiments were performed employing measurements of the relaxation emissions from pre-irradiated Kr samples – unannealed and annealed before exposure to an electron beam. Three emissions were monitored in correlated in real time manner: thermally stimulated luminescence (TSL), thermally stimulated exoelectron emission (TSEE) and total yield of particles via pressure measurements. The energy levels of defects were estimated from the TSL data of the annealed sample. Two types of electron-hole traps created by electronic excitation were identified – close pairs and distant ones. Additional confirmation of the „excited state" mechanism of defect formation was obtained. Analysis of the yields correlation and effect of thermal treatment gave additional arguments in support of the crowdion model of anomalous low temperature post-desorption (ALTpD) from pre-irradiated Kr matrices.

## 1. Introduction

Cryocrystals – crystals which are built from atoms of rare gases or simple molecules, are bound by the weak Van-der- Waals forces and exist only at low temperatures or high pressure. Being the simplest materials convenient for solid state research they attract much attention over the decades. Their properties, especially properties of rare gas solids (RGS) were studied and summarized in a number of books, book chapters and reviews [1-11]. In addition to purely academic interest, RGS found practical application as detectors for "dark matter" search [12] and moderators [13-15]. Free clusters of RGS were used for designing a source of VUV and ultrasoft X-ray radiation [16]. RGS have received the widest application in the field of matrix isolation. The idea to isolate transient, highly reactive molecules in cryogenic environment emerged nearly a hundred years ago and initiated the development of a new field of research – matrix isolation [17, 18]. The high interest and fast development of research based on matrix isolation methods are evidenced by a number of books and reviews [19-29]. If at first the efforts of researchers were focused on the study of molecules placed in a matrix, then at the next stage the focus of research shifted to studying the interaction of embedded molecules with the surrounding matrix and the formation of new compounds with atoms of inert elements [26, 27, 29].

The impressive progress in astronomical observations with a number of space missions has led to an increase in interest and activity in the field of astrochemistry, which includes observations, laboratory simulations, and theoretical simulations. The best demonstration of the field of astrochemistry rapid development is the virtual special issue of J. Phys Chem [30]. Examples of recent publications in the field of matrix isolation are the studies [31-42]. A comprehensive review of recent studies of astrophysical ices by matrix isolation methods is presented in [43]. The review is focused on spectroscopy of astrochemically important molecules, ions and radicals stabilized in cryogenic matrices and experimental modeling of mechanisms of radiation-induced and „in dark" chemical reactions occurring in interstellar, cometary and planetary ices. It should be emphasized that these objects are exposed to various radiation in space (ions, electrons, photons), and when considering the results of laboratory modeling, radiation-induced effects in the matrices themselves should be taken into account. Moreover, solid information about the electronic excitations of the matrix, their self-trapping, and the charge centers formation is needed to consider the effect of energy transfer from the matrix to the dopant on its chemical transformation (see e.g. [37]).

When a fast ion or electron collides with a matrix atom and transfers enough energy to it to overcome the binding forces, the knocked out matrix atom will collide with neighboring atoms, creating a cascade of collisions resulting in the formation of an extended defect and sputtering. Some part of energy deposited by fast incident particle will excite matrix atoms and the electronic excitation energy will relax in a matrix via radiative and nonradiative transitions. The nonradiative transitions are commonly followed by a heat release. However, there is a peculiar type of nonradiative transitions that involves large displacement of a small number of atoms, resulting in the creation of point lattice defects, as well as matrix atom desorption. The inclusion of processes of radiation-induced defect formation is of fundamental importance in elucidating the mechanisms of mass diffusion, desorption and solid-phase chemical reactions of dopants and their fragments. Extensive investigations of the electronically induced defect formation in RGS were performed under excitation with slow electrons and synchrotron radiation [5, 7-9. 11, 44-48]. The basis for the physics of electronically induced lattice rearrangement is a concentration of the excitation energy released in the relaxation process within a volume about that of a unit cell followed by the energy transfer to the surrounding matrix. The spectroscopic investigation of lattice defect creation in the case of exciton self-trapping into molecular states, as occurred in solid Xe, Kr and Ar, were carried out based on an analysis of the luminescence band of molecular type self-trapped excitons (M-STE). This broad band stems from the electronic transitions from the $^{1,3}\Sigma_u^+$ states to a repulsive part of the ground state $^1\Sigma_g$ (indicated as M-band [5, 7, 9]). It has been shown that the M-band in all heavy rare-gas matrices consists of two components – $M_1$ (related to the molecular centers in the defect sites) and $M_2$ (related to the molecular centers formed in the regular lattice at the exciton self-trapping). For the RG matrices with negative electron affinity $E_a$ (Ne and Ar) the so-called „cavity ejection" mechanism of defect formation and desorption operates [5, 7, 9, 49]. It should be noted that the Ar matrix is a borderline case: the mechanisms characteristic of heavier RG matrices also operate in solid Ar. For the RG matrices with positive electron affinity (Xe and Kr) the „cavity ejection" mechanism does not work. The source of energy for the defect production and desorption in these matrices is the energy release in relaxation process. The radiative transition from the excited state of the M-STE to the repulsive part of the ground state results in an appearance of „hot" Rg atoms with an excess kinetic energy of about 0.5 eV sufficient to dislodge some neighboring matrix atom. This mechanism has been called the „ground state" or „excimer dissociation" mechanism [9, 10, 44]. Another mechanism of electronically induced defect formation through self-trapping of an exciton into M-STE states was also proposed – the „excited state" mechanism (see e.g. [44]). According to this mechanism defects are formed during lifetime of the excited molecule $Rg_2^*$. This molecular dimer aligned along the <110> crystallographic directions can be considered as a „dumb-bell'' configuration of the interstitial atom. However, the only stable form of the interstitial atom in the Rg lattice is the split <100> „dumb-bell'' form [50]. Applying the „off-center" concept [5] it was assumed that the short-lived defect of the off-center configuration (dimers shifted along the <110> direction) can be stabilized by the reorientation to the <100> direction. The Frank–Condon transition of this dimer to the ground state will correspond to the transition of the molecular center to the permanent defect level with almost no change in the interatomic distance. The energy needed for the reorientation can be gained from the vibronic system. According to the theory [51], in a system with strong local vibrations the energy is released in a jump-like multiphonon process. The electronically induced defect formation in solid Kr through these mechanisms and the temperature dependence of defect accumulation rate were studied using luminescence method [52, 53].

An effective approach to study radiation induced matrix modification and defect formation is activation spectroscopy – complex of methods, based on investigation of relaxation emissions from preliminary irradiated solids. The most popular method of activation spectroscopy is thermally stimulated luminescence (TSL) [54]. The first measurements of the TSL of RGS were performed after irradiation with electrons [55] and X-rays [56]. Thermally stimulated currents (TSC) of solid Ar undergone synchrotron radiation were detected in [57]. Another current activation technique – thermally stimulated exoelectron emission (TSEE) was employed to probe defects in RG matrices [58]. Due to the high mobility of electrons in RGS [5], TSEE measurements provide information not only on surface-related processes but also on the processes occurring in bulk of the films. Because TSEE and TSL compete with each other, TSEE measurements have been supplemented by TSL recording [59-61]. Information on deep traps in pre-irradiated with an electron beam films was obtained by measuring photon stimulated exoelectron emission (PSEE) [62]. This method has also been used in addition to TSL and TSEE measurements [63, 64]. When studying relaxation processes in RG matrices irradiated with electrons, a new effect was discovered: an anomalously strong low-temperature post-desorption (ALTpD) which was observed at $T \ll T_{sb}$, where $T_{sb}$ is the sublimation temperature [46, 48, 65]. In view of the high sensitivity of TSL, TSEE, and the ejection of atoms from the surface to the sample structure and impurity content, it is obvious that these phenomena must be monitored simultaneously on the same sample. Such an approach was developed and applied to study relaxation processes in the Ar matrix [66].

This study is focused on relaxation processes study in Kr matrices pre-irradiated with an electron beam of subthreshold energy. For such a low-energy electrons the knock-on mechanisms of defect formation and desorption do not work and all radiation effects are driven via electronic excitations. In contrast to the Ar matrix, solid Kr has a positive electron affinity $E_a$ = 0.3 eV [5] and electrons should overcome the barrier to be detected as the TSEE current. However, as our previous study performed on the Xe matrix [48], which has an even higher $E_a$ value of 0.5 eV [5], showed that TSEE currents can be detected (due to the uncompensated negative space charge accumulated at exposure of the sample to an electron beam). All three relaxation emissions were measured simultaneously: TSL, TSEE and thermally stimulated post-desorption (TSpD), that is total yield of particles detected via pressure recording in the sample chamber. Influence of the thermal treatment was elucidated and correlation of the peaks observed was analyzed. An origin of the radiation-induced defects was elucidated and the defect energy levels were estimated. Additional experiments with cycles of irradiation and annealing were performed to trace modification of the relaxation emissions.

## 2. Experimental

### 2.1 Sample preparation and irradiation with an electron beam

Films of solid Kr were grown from the gas phase by deposition onto a metal substrate coated by a thin layer of $MgF_2$, which was cooled down to 7 K in a vacuum chamber with a base pressure of about $10^{-9}$ torr. The high-purity (99.995%) Kr gas was used. The gas-handling system and vacuum chamber were degassed and pumped out before each experiment. The typical deposition rate was kept about $10^{-1}$ μm s$^{-1}$, and the final sample thickness was 50 μm. An open surface of films allows the use of current activation technique, based on exoelectron emission – TSEE, and monitoring the desorption total yield by pressure P measurement both during irradiation and during subsequent heating. The temperature was measured by a calibrated silicon diode sensor mounted directly on the substrate. The programmable temperature controller LTC 60 allowed us to measure and maintain a desired temperature during the sample deposition and irradiation as well as to keep the heating regime. The irradiation with an electron beam was performed in dc regime at 7 K. A tungsten filament served as a source of electrons. An electrostatic lens was used to focus the electrons. The electron beam energy $E_e$ was set to 500 eV with the current density of 30 μA cm$^{-2}$. We used slow electrons to avoid the knock-on defect formation sputtering. The beam covered a sample area of 1 cm$^2$. The sample heating under electron beam did not exceed 0.1 K. The radiation dose was varied by an exposure time. The effect of annealing at 40 K on relaxation emissions of pre-irradiated Kr samples was studied.

**2.2 Detection of relaxation emissions: electrons, photons and particles**

Relaxation processes were stimulated by heating of pre-irradiated samples with a constant rate of 3.2 K min$^{-1}$. Measurements were performed in the temperature range of 7 – 40 K. Three relaxation emissions were monitored in correlated manner: electrons, photons and desorbing particles. Being promoted to the conduction band by heating electrons detrapped from shallow traps either neutralize positively charged species yielding TSL photons or escape from the sample yielding TSEE. Stimulated currents were detected with an electrode kept at a small positive potential $V_F$=+9 V and connected to the current amplifier FEMTO DLPCA 200. The stimulated luminescence of solid Kr appears as the result of recombination of self-trapped hole $Kr_2^+$ with detrapped electron yielding self-trapped exciton STE of molecular configuration $Kr_2^*$. Its radiative decay, the well-known M-band [5, 7-9] situated in the VUV range at 8.5 eV. The VUV photons were detected by a Hamamatsu R5070 photo-multiplier tube attached to the UV window covered by a thin film of $C_7H_5NaO_3$ used as a sensitizer. The total yield of desorbing particles – thermally stimulated post desorption (TSpD) was monitored using a Compact BA Pressure Gauge PBR 260 calibrated with a flow rate controller. For the comparison purpose the total yield of particles from unirradiated sample during its heating, so-called temperature programmed desorption (TPD), also was detected. Special experiments with cycles of irradiation and annealing were performed to probe defect generation and matrix reaction. In these experiments we have stopped the controlled annealing at a temperature well below where the sample loss occurs, and irradiated the sample again after re-cooling it to the 7 K. Subsequently this cycle (irradiation by electrons for 30 min; heating to 40 K, while recording the TSL, TSEE and pressure signals; annealing at this temperature for 5 min and re-cooling then back to 7 K) was repeated up to five times. The entire control of the experiment and the simultaneous acquisition of the TSEE and TSL yields along with the vacuum chamber pressure measurements as well as temperature control were accomplished with the help of a program written for these experiments.

3. Results and discussion

In TSL spectra the M-band appears as a result of the recombination reaction of self-trapped or trapped holes (STH/TH) with thermally detrapped electrons:

$$Kr_2^+ + e^- \rightarrow Kr_2^* + \Delta E_1 \quad (1)$$

$$Kr_2^* \rightarrow Kr + Kr + \Delta E_2 + h\nu \quad (2),$$

where $h\nu = 8.5$ eV. Both stages of this phenomenon are accompanied by energy release $\Delta E_1$ and $\Delta E_2$. The glow curves for this emission, taken from annealed and unannealed samples of solid Kr pre-irradiated with a 500 eV electron beam are presented in Figure 1.

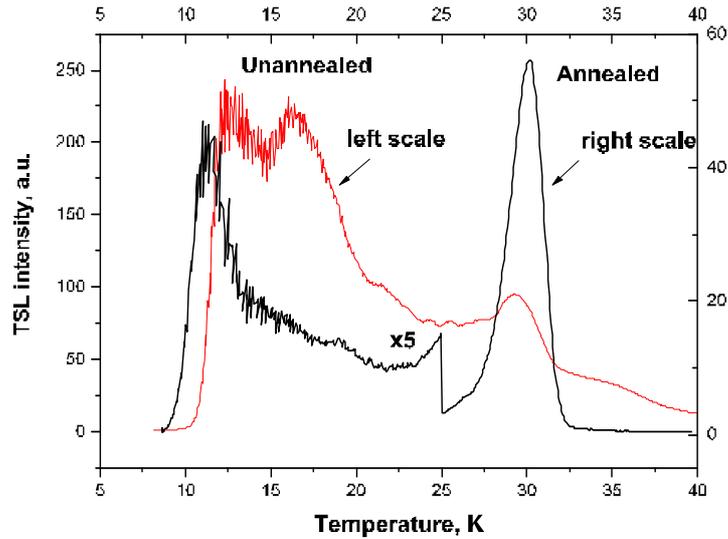

Fig. 1. The VUV TSL curves taken from unannealed and annealed at 40 K films of Kr. Irradiation was performed with a 500 eV electron beam at 7 K during 30 min.

The glow curve for unannealed film shows, in fact, continuous distribution of traps in whole temperature range up to 40 K with three pronounced peaks at 12.5, 16 and 29 K. Note, that the heating range was chosen in such a way to avoid sample evaporation ($T_{sb}$ =45 K). Annealing strongly suppresses and simplifies the TSL curve, which shows only two peaks: a weak peak at about 11.3 K and a strong one at 30 K. The intensity of the main peak at 30 K only slightly decreases in comparison with that measured on an unannealed sample, evidencing its relation mainly to radiation-induced defects. However, the shift of this peak by 1 K points to some contribution of the growth defects modifying the ascending part of the peak. Strong suppression of the low-temperature peak which appears to be ten times weaker in an annealed sample, points to its connection mainly with growth defects in the surface/subsurface layers. The shift of the peak towards lower temperature by 1 K upon annealing indicates appearance of more shallow traps. Changes in the TSEE yield as a result of annealing are presented in Figure 2.

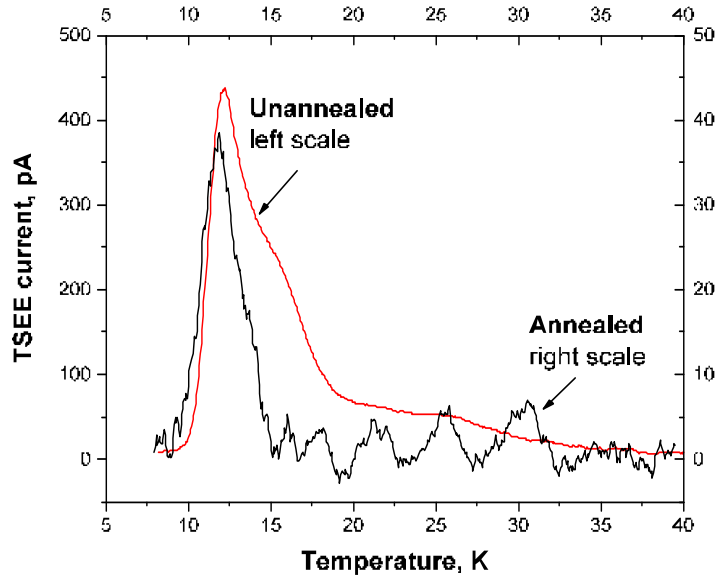

Fig. 2. The TSEE yields taken from unanneaed and annealed at 40 K films of Kr. Irradiation was performed with a 500 eV electron beam.

A distinctive feature of the TSEE yields of both types of samples is the absence of a pronounced feature at 30 K in contrast to the TSL yields. It is worth noting that TSL and TSEE are competitive processes. On the one hand, both of them are triggered by the release of electrons. On the other hand, the electrons leaving the sample do not participate in the recombination reaction. The recombination probability depends on configuration of STH/TH-electron pair. For close pairs the recombination probability is 1 and they do not contribute to the TSEE yield. The nearly zero TSEE yield around 30 K indicates connection of the 30 K maximum in the TSL with radiation-induced interstitial - vacancy (i-v) close pairs. The low-temperature feature at 12 K in the TSEE yield demonstrates a similar behavior with that in the TSL yield – a tenfold suppression, in the annealed sample, which confirms its identification as associated mainly with growth defects. Note, that the FWHM (full width at half maximum) of the TSEE peak recorded for the annealed sample is two times narrower than that for the unannealed one, which suggests that some small part of the surface/subsurface defects can be formed under the action of irradiation.

A phenomenon closely related to relaxation processes in surface/subsurface layers is anomalously strong low-temperature post-desorption – ALTpD. First this effect was detected in solid Kr [46] while studying properties of low-temperature unannealed condensates. The phenomenon was interpreted as a consequence of intrinsic charge recombination resulting in „excimer dissociation" via radiative transition of $Kr_2^*$ to a repulsive part of the ground state potential curve (the transitions $^{1,3}\Sigma_u^+ \to {}^1\Sigma_g$). We extended these measurements and elucidated the influence of annealing on the phenomenon. Figure 3 shows the temperature dependence of the total yields of particles monitored via pressure records from unannealed and preliminary annealed before irradiation Kr films.

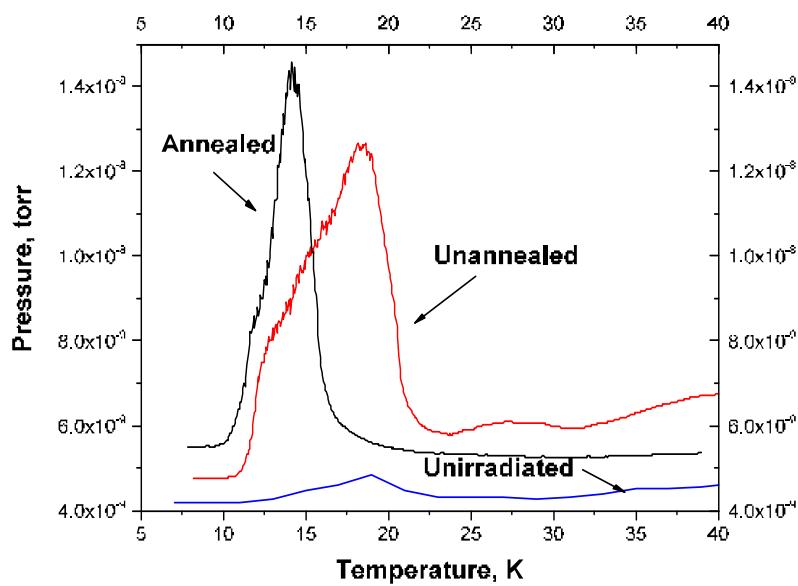

Fig. 3. The pressure in the chamber recorded during heating of pre-irradiated films, both unannealed and annealed before irradiation with a 500 eV electrons. The TPD curve of unirradiated sample is shown for comparison.

As can be seen, for both types of samples, anomalously strong pressure rise was observed in the range of temperatures much lower than the sublimation temperature $T_{sb}$ of solid Kr. It appears that the annealing of sample drastically changes the ALTpD yield. The pressure curve became twice narrower, its maximum shifted from 18 K to 15 K and increased in intensity. In order to be sure that the effect is caused by irradiation the yield of particles from the unirradiated and unannealed sample was measured using the same heating mode. The resulting curve of the so-called temperature programmed desorption (TPD) is shown in Fig. 3. Some weak increase in pressure was observed with a maximum at 18 K for the unirradiated sample. The appearance of this feature can be interpreted as desorption induced by growth defects due to the weakening of binding forces at the defect sites. A stronger increase in pressure in the low-temperature range of 11–21 K upon heating of pre-annealed and pre-irradiated sample confirms the radiative origin of the phenomenon.

To get more close insight into the radiation-induced relaxation processes, let's analyze all three simultaneously measured relaxation emissions together. Figures 4 and 5 show them for unannealed (Fig. 4) and annealed before irradiation (Fig. 5) samples.

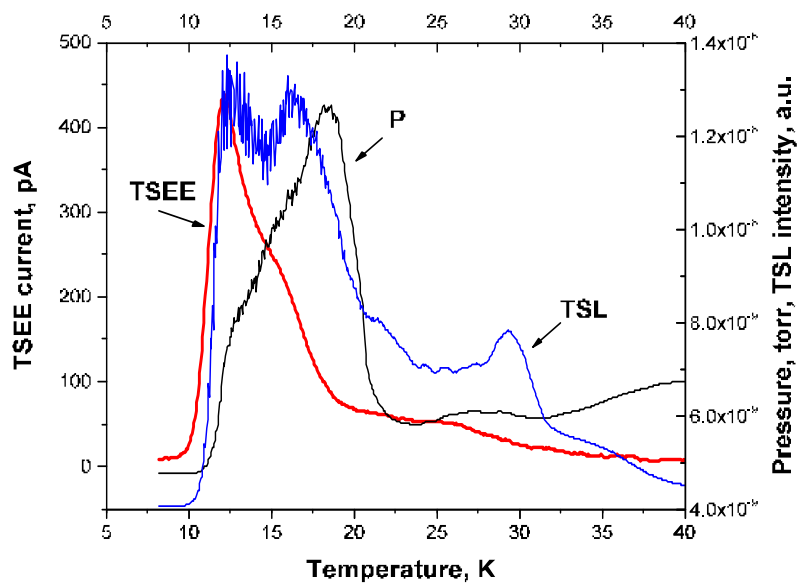

Fig.4. Yields of the TSL in VUV range, TSEE and pressure curve (P) measured for the unannealed sample at linear heating.

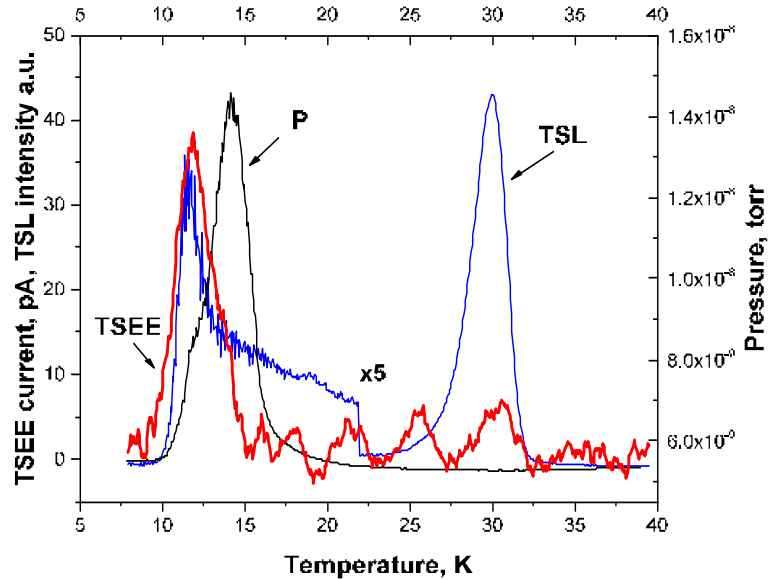

Fig. 5. Yields of the TSL in VUV range, TSEE and pressure curve (P) measured for the annealed sample at linear heating.

As it was mentioned, because of positive electron affinity ($E_a$ = 0.3 eV) of solid Kr [5] there exists a barrier for electrons to escape from the surface of the sample. So one could expect some shift of the TSEE peak position to the higher temperatures as compared to that in VUV TSL. However, as can be seen from Figs. 4 and 5 there is no such shift. The coincidence of the first peaks of the TSL and TSEE yields, despite the presence of a barrier for electron to escape, is due to the negative space charge accumulated in traps upon exposure to an electron beam. Additional evidence of the negative charge accumulation is the observation of electron „afteremission". This electron emission was found in unannealed and weaker in annealed Kr samples after the electron beam was switched off. The electric field created by this space charge facilitates overcoming the energy barrier, caused by the positive $E_a$ of solid Kr, and the escape of electrons from the sample. This „afteremission" decayed exponentially, and measurements of thermally stimulated emissions (TSL, TSEE and yield of particles) were started when it became practically zero. The absence of a peak at 30 K in the yield of TSEE from both kinds of samples, while it dominates in the yield of TSL of annealed samples, indicates its radiation origin and the connection of this peak with close (STH/TH-e) pairs. In view of high kinetic energy of Kr atoms after dissociation (0.485 eV each) creation of close pairs via „excimer dissociation" mechanism looks less probable. Taking into account that the distance between the „dumb-bell" atoms in Kr is 0.339 nm [67], which is only slightly larger than $r_e$ of the $Kr_2^+$ molecular ion ($r_e$ = 0.28 nm [68]), one can assume that the positive charge – trapped hole (TH), is localized at the interstitial of „dumb-bell" configuration while the electron is localized at the vacancy. The atomic fraction of equilibrium vacancies is extremely low <$10^{-9}$ at low temperatures and only nonequilibrium vacancies are involved in the process. The close intensities of the peak at 30 K for the annealed and unannealed samples (as seen in Fig. 1) indicate that the mechanism for the formation of these pairs does not require the presence of a long-range order. The data obtained supports the „excited state" mechanism of defect formation. Its two-stage character, i.e., the shift of the excimer along crystallographic directions of the <110> type with its subsequent reorientation to the <100> direction, takes some time; therefore, only the triplet excited state $^3\Sigma_u^+$ of the excimer with longer lifetime ($\tau_3$ = 3200 ns [5]) can control the process. No permanent defect can be formed by the „excited state" mechanism through the singlet excited state $^1\Sigma_u^+$ with a rather short lifetime $\tau_1$ = 1.2 ns [5]. However, the radiative annihilation of the singlet state, accompanied by the appearance of „hot" atoms, can generate point defects in the lattice at the second stage (2) of the recombination process. It should be mentioned that upon exposure to an electron beam creation of excited ionic molecular centers $Kr_2^{+*}$ is possible. According to [69], the lowest excited state of $Kr_2^{+*}$ – $I(3/2)_g$, has a minimum at $r_e$ = 0.31-0.32 nm, which coincides with the atom separation in „dumb-bell" configuration. The involvement of $Kr_2^{+*}$ centers in relaxation processes in pre-irradiated Kr matrices requires further study. Based on the measured TSL curve of the annealed sample we estimated the trap depth energy $E_d$ corresponding to the 30 K TSL peak. The descending part of the TSL peak free of overlapping bands, was used for estimation of the trap depth energy: $E_d = kT_m^2/(T_2-T_m)$, where $T_m$ is the temperature at the band maximum, $T_2$ – the temperature on the descending part of the curve at half the height of the peak. Using the half-width method [54], and assuming the first-order kinetics we obtained $E_d$ = 77.4 meV. This value is higher than the value $E_d$ = 43 meV obtained in [56] for the maximum at 30 K in TSL of a bulk Kr crystal after 2-hour X-ray irradiation at 15 K. Note that in our measurements, extended to 55 K, we did not observe any traces of the dominant TSL peak at 40 K, detected after X-ray irradiation [56]. Estimating the low-temperature trap depth energy by the ascending part of the TSL curve we obtained $E_d = 1.51 k T_m T_1/(T_m - T_1) = 20.3$ meV (here $T_1$ is the temperature on the ascending part of the curve at half the height of the peak). The thresholds for the TSL, TSEE and the pressure rise from preliminary annealed sample coincide, as shown in Fig. 5. The shoulder of the first TSL peak of the annealed sample extended to higher temperatures indicates the presence of other traps, and the shift of the pressure maximum relative to the first TSL maximum indicates the participation of these traps in the low-temperature burst of particles from pre-irradiated Kr film. The release of electrons from these traps (including the first one), followed by their recombination with $Kr_2^+$ centers, is the stimulating factor for triggering the ALTpD phenomenon. It is interesting to compare the pressure curve with the TSL curve for the unannealed sample. The multipeak TSL curve of such sample fitted by 5 Gaussian peaks is shown in Fig. 6.

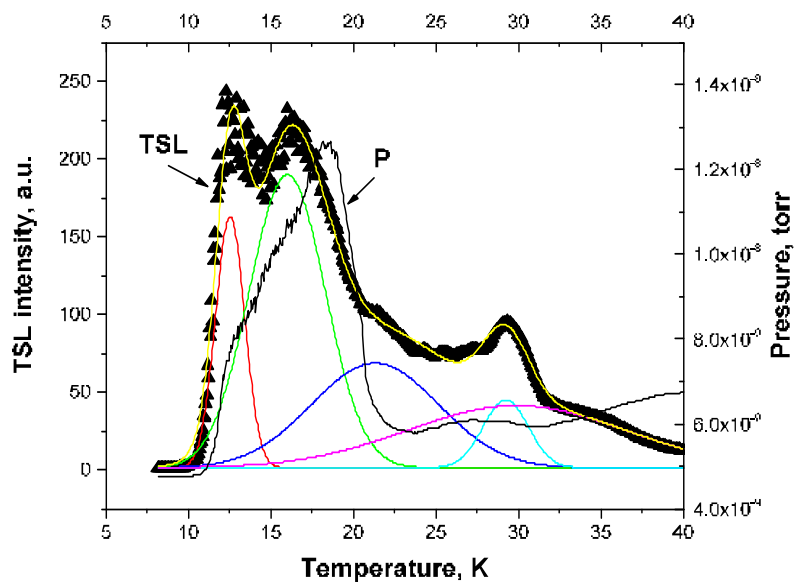

Fig. 6. The TSL curve of the unannealed sample, fitted with 5 Gaussian peaks, and recorded simultaneously the total particle yield (P) from the pre-irradiated sample.

As can be seen, the pressure curve almost follows the behavior of peak 2 (green curve) with some contribution of the other low-temperature peaks. But there is no contribution from the trap related to the 30 K TSL maximum. Obviously, the positions of the $Kr_2^+$ recombination centers relative to the surface determine their contribution to the ALTpD phenomenon. This poses a question of energy transfer.

Let's discuss the applicability of the crowdion model to explain this phenomenon. A crowdion is a chain of atoms compressed due to the presence of an extra atom in the row. In the RGS with FCC lattice the close-packed atomic rows are oriented along crystallographic directions of the type <110>. As shown by theoretical calculations [70], crowdions exist in the lattices of solidified Ar and Kr. In this work, the numerical values of the main parameters of crowdions were obtained: the self-energy $E_s$, the effective mass $m_s$, and the characteristic length $\lambda_s$ for Ar and Kr FCC matrices. The crowdion energy $E_s$ in Kr ($E_s = 0.42$ eV) is appeared to be comparable with the energy $E_a$ transferred to the ground state Kr atoms ($E_a = 0.485$) eV after dissociation of neutralized $Kr_2^+$ center at the radiative electronic transition to the ground state. The characteristic crowdion width in solid Kr, according the numerical estimates [70], is several times greater than the parameter of a close-packed row, viz. $\lambda_s = 2.76b$ (b is the translations vector along the atomic row). Using the Kr lattice parameters given in Ref. [5], we obtain an estimate of the characteristic size of the crowdion nucleus in a three-dimensional Kr crystal $\lambda_s = 2.76b = 1.1$ nm. The effective mass of the crowdion in Kr is rather small $m_s = 0.3m_a$ ($m_a$ is the atomic mass) indicating its high mobility. The movement of a crowdion in a discrete chain of atoms is connected with overcoming the potential barriers $\Phi_m = 0.17$ eV between neighboring energy minima and the crowdion can overcome them by quantum tunneling. The ability of a crowdion as a dynamic defect that can easily move along a close-packed row of atoms makes the crowdion model the most adequate for interpreting the ALTpD phenomenon. A strong argument in favor of this model is the increase in the pressure rise in the annealed sample, while the TSL yield, which serves as a recombination marker, is strongly suppressed. Lattice ordering during annealing increases the length of ordered close-packed rows of atoms and facilitates the transfer of energy from deeper layers to the matrix surface. Consideration of the crowdion parameters in the Ar and Xe matrices shows that the crowdion model can also be used to explain the ALTpD phenomenon in the matrices of other rare gases, Ar and Xe. Note that the data obtained in [70] on the parameters of the crowdion in Ar and Kr can be extended to Xe, taking into account the law of the corresponding states. During the recombination of $Rg_2^+$ centers in the depth of the sample, the formation of an „extended" defect, a crowdion, is also possible. It should be underlined that the crowdions in RG matrices are of importance considering the electronically induced desorption of matrix atoms as well as diffusion processes.

We performed additional experiments with cycles of irradiation and heating to trace modification of the relaxation emissions. The Kr sample was first deposited at 7 K, then annealed at 40 K for 5 min, re-cooled back to 7 K, and subsequently exposed to a 500 eV electron beam for 30 min. After that it was heated up to 40 K while recording the TSL, TSEE and pressure signals, and then re-cooled back to 7 K. Subsequently this cycle was repeated up to five times keeping the same parameters of the beam and the heating mode. In these experiments the controlled heating after each cycle of irradiation was stopped at 40 K, which is below the sample loss temperature. As cycling proceeded, further annealing of defects occurred, while defects induced by irradiation, both structural and charge ones ($Kr_2^+$ and electrons), were restored in each irradiation cycle. Of course, the sample thickness somewhat decreased during cycling due to the electron-induced desorption of atoms under the beam, but this effect did not affect the results, taking into account the significant initial film thickness (50 μm). The behavior of TSL during cycling is shown in Fig. 7.

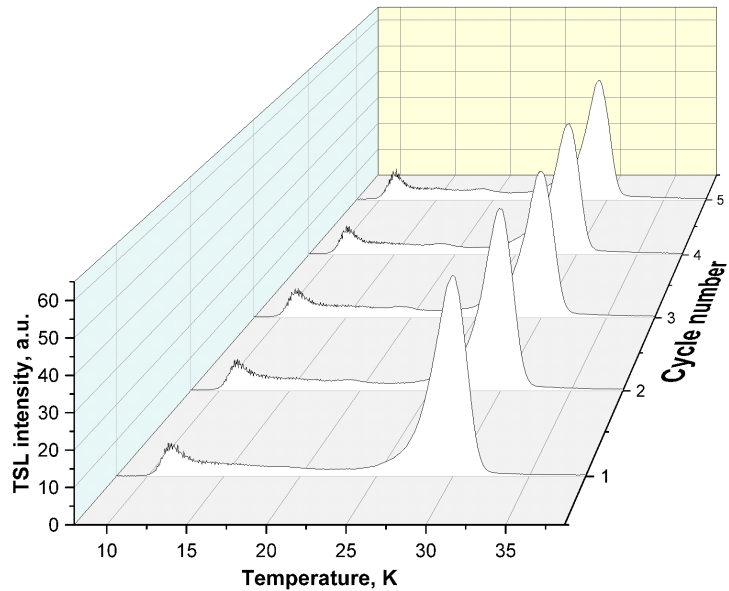

Fig. 7. Three-dimensional plot of the TSL yield for five cycles of irradiation and annealing.

As can be seen, strong TSL reappeared in each cycle, only slightly modified. A comparison of the TSL yields measured in the first and fifth cycles is shown in Fig. 8.

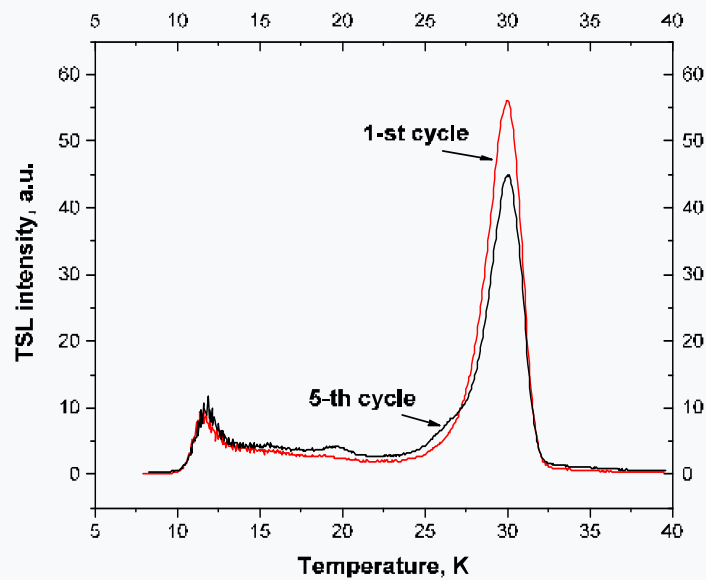

Fig. 8. TSL glow curves taken after the first and fifth cycles of exposure to an electron beam.

The intensity of the TSL maximum at 30 K decreased by a factor of 1.2 during cycling, and its ascending part slightly increased in the wing, which indicates a certain contribution of growth defects to the 30 K band. As can be seen, only the descending part of the 30 K maximum has not changed. The high-temperature maximum did not change its position while the low-temperature one shifted very slightly (by 0.2 K) towards higher temperatures after 5 irradiation cycles. For the low-temperature TSL peak the ascending part of the curve remained unchanged, while the descending part slightly increased. It should be noted that it is the unchanged parts of the peaks in the TSL curve that were used to estimate the defect energies $E_d$ above. The TSL intensity somewhat increased in the range of 12–27 K, and extremely weak bands started to emerge at 16 and 20 K, which were present in the TSL of the unannealed sample (see Fig. 6). This effect of heating cycles is believed to be due to thermal diffusion of defects followed by the formation of more complex ones.

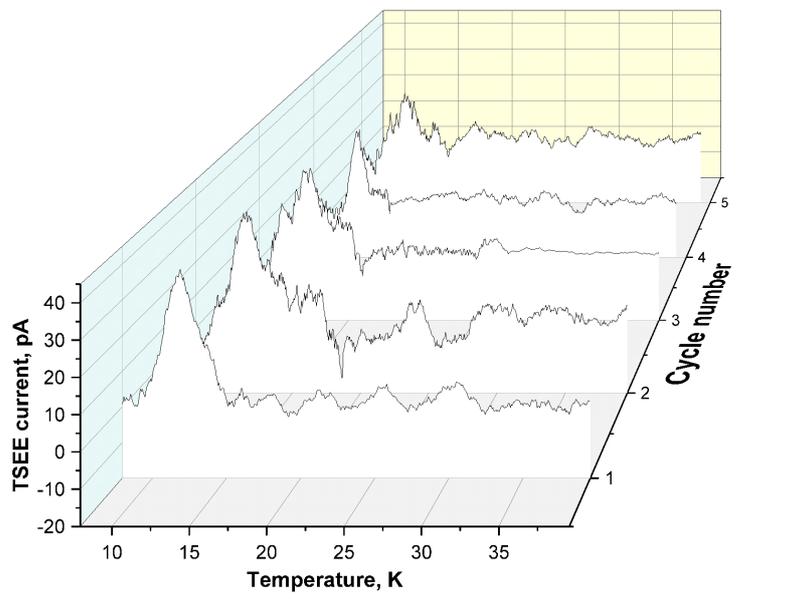

Fig. 9 The TSEE yield modification with cycles of irradiation and heating.

The TSEE yield was very low since cycling started with irradiation of the annealed sample (see Fig. 2) and decreased from cycle to cycle as shown in Fig. 9. This trend contrasts with the behavior of the low-temperature feature of TSL, which, taking into account the competition between TSEE and TSL, suggests an increasing role of the recombination process $Kr_2^+ + e^- \rightarrow Kr_2^* + \Delta E_1$, followed by the production of „hot" Kr atoms in reaction (2), stimulating ALTpD from the surface.

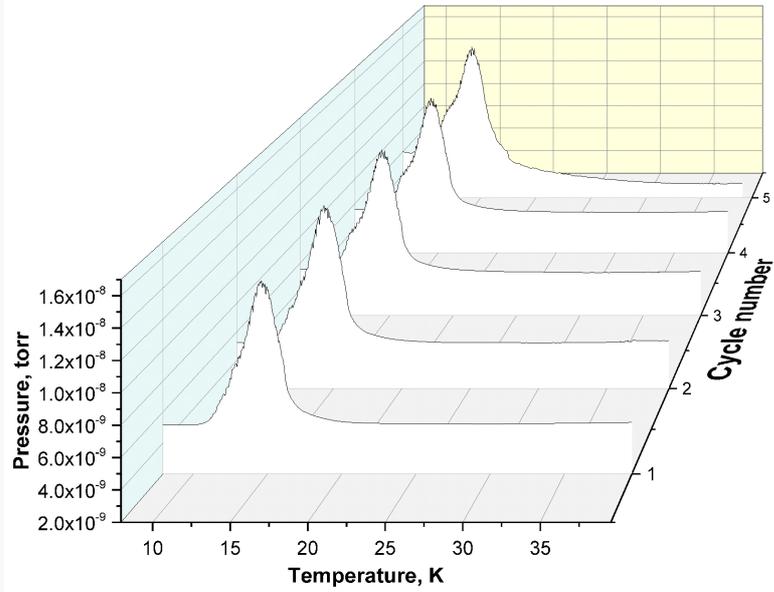

Fig. 10. The pressure rise modification with cycles of irradiation and heating.

Fig. 10 shows the ALTpD behavior with cycles of exposure to an electron beam and heating. The corresponding curves recorded during the first and last cycles are presented in Fig. 11.

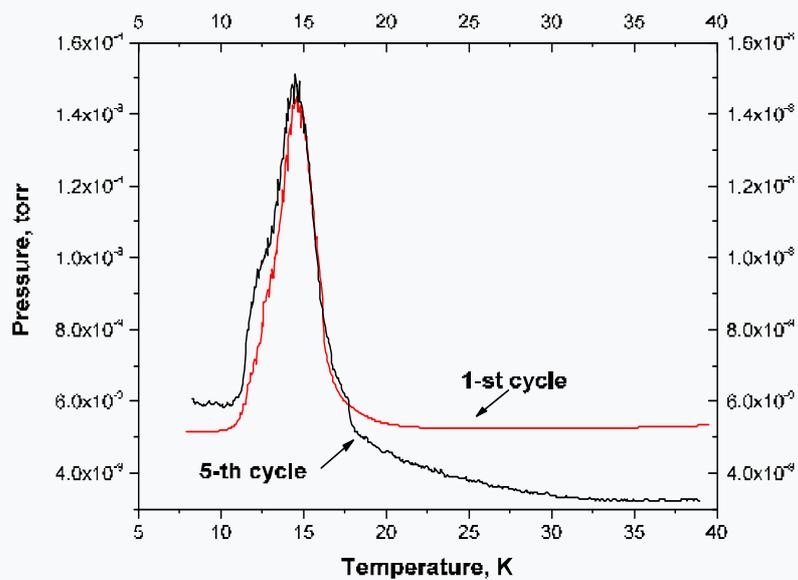

Fig. 11. The pressure peaks upon the first and fifth cycles of annealing pre-irradiated sample.

The pressure peak changed very little. The initial pressure slightly increased in each cycle, but the position of the pressure maximum remained constant within the measurement accuracy. The half-width of pressure peak increased with cycling, and its shape became asymmetric, which indicates the inclusion of new traps in the process. Significantly, the difference between the initial pressure and its maximum value remained almost constant. When considering the ALTpD effect, it should be taken into account that all measurements in this study were carried out in the dynamic pumping out mode, i.e. the increase in pressure was actually much greater. The effect of covering the irradiated sample with a layer of non-irradiated Kr film was checked before recording the thermally stimulated desorption. It turned out that a thin film (~10 nm) of unirradiated Kr, condensed on top of the irradiated sample before heating, only slightly suppressed the pressure rise, demonstrating efficient energy transfer through a dozen atomic layers.

**Summary**

The study was focused on relaxation phenomena in Kr matrices exposed to an electron beam. The subthreshold energy electrons were used to exclude the knock-on defect formation and desorption. Two types of samples were used – quench condensed films and films annealed at 40 K before irradiation. The experiments were performed employing methods of activation spectroscopy – thermally stimulated luminescence (TSL), thermally stimulated exoelectron emission (TSEE) and monitoring of the total yield of particles via pressure recording. Measurements of the relaxation emissions were carried out in correlated in real time manner from the same sample. Additional experiments with cycles of irradiation and heating to trace modification of the relaxation emissions were performed. The Kr sample was first deposited at 7 K, then annealed at 40 K for 5 min, re-cooled back to 7 K, and subsequently exposed to a 500 eV electron beam for 30 min. After that it was heated up to 40

K with a constant rate of 3.2 K min$^{-1}$ while recording the TSL, TSEE and pressure signals, and then re-cooled back to 7 K. This cycle was repeated up to five times. Analysis of the data obtained revealed two types of electron-hole traps created by electronic excitation – close pairs and distant ones, and provided support for the „excited state" mechanism of defect formation. Comparison of the yields correlation and effect of annealing gave additional arguments in support of the crowdion model of anomalous low temperature post desorption (ALTpD) from pre-irradiated Kr matrices.

**References**

1. M.L. Klein, J.A. Venables, *Rare gas solids*; Academic Press, New York (1976).
2. M.L. Klein, J.A. Venables, *Rare gas solids*; Academic Press, New York (1977).
3. *Cryocrystals*, B.I. Verkin and A.F. Prikhodko (eds.), Naukova Dumka, Kiev (1983).
4. N. Schwentner, E.-E. Koch, J. Jortner, *Electronic excitations in Condensed rare gases*; Springer-Verlag: Berlin (1985).
5. K.S. Song and R.T. Williams, *Self-Trapped Excitons*, Springer-Verlag, Berlin (1996).
6. *The Physics of Cryocrystals*, V.G. Manzhelii, Yu.A. Freiman, M.L. Klein, and A.A. Maradudin (eds.), AIP Publisher, Woodbury, MA (1997).
7. G. Zimmerer, *Creation, motion and decay of excitons in rare-gas solids*, in: U. M. Grassano, N. Terzi (eds.), *Excited-State Spectroscopy in Solids*, North-Holland, Amsterdam (1987).
8. I.Ya. Fugol, *Adv. Phys*. **27**, 1 (1978).
9. I.Ya. Fugol, *Adv. Phys*. **37**, 1 (1988).
10. R.E. Johnson and J. Schow, *Mat. Fys. Medd. K. Dan. Vidensk. Selsk*., **43**, 403 (1993).
11. E.V. Savchenko and Yu.A. Dmitriev, *New Aspects of Relaxation Processes in Cryogenic Solids,* in *Applied Physics in the 21st Century* (Horizons in World Physics. Volume 269) P. Raymond (ed.) Valencia, Nova Science Publishers, New York, (2010) p. 113-162.
12. M. Guarise**,** *Eur. Phys. J. Plus* **137,** 673 (2022); https://doi.org/10.1140/epjp/s13360-022-02876-4
13. B. Shizgal and D. R. A. McMahon, Phys. Rev. A 32, 3669–3680 (1985).
14. G.R. Massoumi, N. Hozhabri, W.N. Lennard, P.J. Schultz, S.F. Baert, H.H. Jorch, and A.H. Weiss, *Rev. Sci. Instrum.* **62**, 1460 (1991); doi:10.1063/1.1142467
15. E. Morenzoni, M. Birke a, H. Glückler, A. Hofer, J. Litterst, M. Meyberg, C. Niedermayer, Th. Prokscha, G. Schatz and Th. Wutzke, *Hyperfine Interactions***, 106**, 229 (1997).
16. V.L. Vakula, O.G. Danylchenko, Yu.S. Doronin, G.V. Kamarchuk, O.P. Konotop, V.N. Samovarov, A.A. Tkachenko, *Fiz. Nizk. Temp.* **46**, 180 (2020)/*Low Temp. Phys.* **46**, 145 (2020), an information on the source is available via http://ilt.kharkiv.ua/bvi/technology/otd13/gjs_e.html
17. L. Vegard, *Nature*, **114**, 357 (1924).
18. E. Whittle, D. A. Dows and G. C. Pimentel, *J. Chem. Phys*. **22**, 1943 (1954).
19. B. Meyer, *Low Temperature Spectroscopy*: *Optical Properties of Molecules in Matrices*, *Mixed Crystals*, *and Frozen Solutions*, American Elsevier, New York (1971).
20. M. Moskovits and G. A. Ozin, *Cryochemistry*, Wiley, New York (1976.)

21. A.J. Barnes, *Matrix Isolation Spectroscopy*, NATO ASI Series, D, Reidel Publishing Company, Dordrecht (1981).
22. *Chemistry and Physics of Matrix-Isolated Species*, L. Andrews and M. Moskovits (eds.) Elsevier, Amsterdam (1989).
23. V.E. Bondybey, A.M. Smith, and J. Agreiter, *Chem. Rev.* **96**, 2113 (1996); https://doi.org/10.1021/cr940262h
24. V.E. Bondybey, M. Räsänen, and A. Lammers, *Ann. Rep. Prog. Chem.*, Sect. C: *Phys. Chem.* **95**, 331 (1999).
25. V.A. Apkarian, N. Schwentner, *Chem. Rev.* **99** 1481 (1999).
26. M. Pettersson, J. Lundell, and M. Räsänen, *Eur. J. Inorg. Chem.* **1**, 729 (1999).
27. A.V. Nemukhin, L.Yu. Khriachtchev, B.L. Grigorenko, A.V. Bochenkova and M. Räsänen, *Russ. Chem. Rev.* **76** 1085 (2007).
28. *Physics and Chemistry at Low Temperatures,* L. Khriachtchev (ed.), Pan Stanford Publishing, Singapure (2011).
29. W. Grochala, M. Räsänen, *Noble-Gas Chemistry,* Ch. 13 in *Physics and Chemistry at Low Temperatures,* L. Khriachtchev (ed.) Pan Stanford Publishing, Singapure (2011) p.447.
30. R.C. Fortenberry, R.J. McMahon, and R.I. Kaiser, *J. Phys. Chem. A*, **126**, 6571 (2022).
31. A.D. Volosatova, D.A. Tyurin, V.I. Feldman, *J. Phys. Chem. A*, **126**, 3893 (2022).
32. E. Savchenko, I. Khyzhniy, S. Uyutnov, M. Bludov, V. Bondybey, *Nucl. Instr. Meth. B*, **536** 113 (2023).
33. I.S. Sosulin, D.A. Tyurin and V.I. Feldman, *J. Chem. Phys.* **154**, 104310 (2021); https://doi.org/10.1063/5.0041159
34. T. Putaud, C. Wespiser, M. Bertin, J.-H. Fillion, Y. Kalugina, P. Jeseck, A. Milpanis, L. Philippe, P. Soulard, B. Tremblay, C. Tuloup, P. Ayotte and X. Michaut, *J. Chem. Phys.* **156**, 074305 (2022); https://doi.org/10.1063/5.0079566
35. Yu.S. Doronin, V.L. Vakula, G.V. Kamarchuk, A.A. Tkachenko, I.V. Khyzhniy, S.A. Uyutnov, M.A. Bludov, E.V. Savchenko, *Fiz. Nizk. Temp.* **47,** 1157 (2021)/*Low Temp. Phys.* **47,** 1058 (2021).
36. J.A. Tan and J.-L. Kuo, *J. Chem. Phys.* **154**, 134302 (2021); https://doi.org/10.1063/5.0044703
37. E.V. Savchenko, I.V. Khyzhniy, S.A. Uyutnov, M.A. Bludov**,** and V.E. Bondybey, *J. Mol. Structure*, **1221,** 128803 (2020), https://doi.org/10.1016/j.molstruc.2020.128803
38. P.K. Sruthi, Swaroop Chandra, N. Ramanathan and K. Sundararajan, *J. Chem. Phys.* **153**, 174305 (2020); https://doi.org/10.1063/5.0031162
39. E. Savchenko, I. Khyzhniy, S. Uyutnov, M. Bludov, and V. Bondybey, *Nucl. Instr. Meth. B*, **469,** 37 (2020), DOI: 10.1016/j.nimb.2020.02.031
40. I.V. Khyzhniy, S.A. Uyutnov, M.A. Bludov**,** E.V. Savchenko and V.E. Bondybey, *Fiz. Nizk. Temp.* **45**, 843 (2019)/*Low Temp. Phys.* **45**, 721 (2019), DOI: 10.1063/1.5111295
41. T. Wakabayashi, T. Momose and M.E. Fajardo, *J. Chem. Phys.* **151**, 234301 (2019); https://doi.org/10.1063/1.5134454
42. N.N. Kleshchina*,* I.S. Kalinina*,* I.V. Leibin*,* D.S. Bezrukov, and A.A. Buchachenko, *J. Chem. Phys.* **151**, 121104 (2019); https://doi.org/10.1063/1.5118876
43. V.I. Feldman S.V. Ryazantsev and S.V. Kameneva, *Russ. Chem. Rev.* **90,** 1142 (2021); DOI: 10.1070/RCR4995
44. E. Savchenko, A. Ogurtsov, I. Khyzhniy, G. Stryganyuk, G. Zimmerer, *Phys. Chem. Chem. Phys.* **7,** 785 (2005).
45. A.N. Ogurtsov, E.V. Savchenko, S. Vielhauer, G. Zimmerer, *J. Luminescence*, **112,** 97 (2005).
46. E.V. Savchenko , I.V. Khyzhniy, S.A. Uyutnov, G.B. Gumenchuk, A.N. Ponomaryov, G. Zimmerer, V.E. Bondybey, *Nucl. Instr. Meth. B*. **267,** 1733 (2009); doi:10.1016/j.nimb.2009.01.155.

47. E.V. Savchenko, G. Zimmerer, V.E. Bondybey, *J. Luminescence*, **129,** 1866 (2009); doi:10.1016/j.jlumin.2009.01.040.
48. E.V. Savchenko, I.V. Khyzhniy, S.A. Uyutnov, G.B. Gumenchuk, A.N. Ponomaryov, V.E. Bondybey, *Nucl. Instr. Meth. B* **268,** 3239 (2010).
49. E.V. Savchenko, A.N. Ogurtsov, and G. Zimmerer, *Fiz. Nizk. Temp*. **29,** 356 (2003)/*Low Temp. Phys.* **29,** 270 (2003).
50. M. Doyama and R.M.J. Cotterill, *Phys. Rev.* **1**, 832 (1970).
51. V. Hizhnyakov, *Phys. Rev. B* **53,** 13981 (1996).
52. I.Ya. Fugol, E.V. Savchenko, A.N. Ogurtsov, O.N. Grigorashchenko, *Physica B,* **190,** 347 (1993).
53. E.V. Savchenko, A.N. Ogurtsov, O.N. Grigorashchenko, S.A. Gubin, *Chemical Physics,* **189,** 415 (1994).
54. D.R. Vij, *Thermoluminescence,* in*: Luminescence of Solids*, D.R. Vij (ed.), Plenum Press, New York, (1998), pp. 271-308.
55. A.N. Ogurtsov, E.V. Savchenko, O.N. Grigorashchenko, S.A. Gubin and I.Ya. Fugol', *Fiz. Nizk. Temp.* **22,** 1210 (1996)/*Low Temp. Phys*., **22**, 926 (1996).
56. M. Kink, R. Kink, V. Kisand, J. Maksimov and M. Selg, *Nucl. Instr. Meth. B*, **122**, 668 (1997).
57. A. Schrimpf, C. Boekstiegel, H.-J. Stökman, T. Bornemann, K. Ibbeken, J. Kraft and B. Herkert, *J. Phys.: Condens. Matter*, **8**, 3677 (1996).
58. M. Frankowski, E.V. Savchenko, A.M. Smith-Gicklhorn, O.N. Grigorashchenko, G.B. Gumenchuk, and V.E. Bondybey, *J. Chem. Phys*. **121,** 1474 (2004).
59. E.V. Savchenko, O.N. Grigorashchenko, A.N. Ogurtsov, V.V. Rudenkov, G.B. Gumenchuk, M. Lorenz, A.M. Smith-Gicklhorn, M. Frankowski, V.E. Bondybey, *Surface Review and Letters,* **9,** 353 (2002).
60. E.V. Savchenko and V.E. Bondybey, *Phys. Status Solidi A*, **202**, 221 (2005).
61. E.V. Savchenko, I.V. Khyzhniy, S.A. Uyutnov, G.B. Gumenchuk, A.N. Ponomaryov, and V.E. Bondybey, *Nucl. Instr. Meth. B,* **277,** 131 (2012); doi:10.1016/j.nimb.2011.12.042.
62. G.B. Gumenchuk, A.N. Ponomaryov, A.G. Belov, E.V. Savchenko, and V.E. Bondybey, *Fiz. Nizk. Temp*. **33,** 694 (2007)/*Low Temp. Phys*. **33,** 523 (2007); DOI:10.1063/1.2746243
63. E.V. Savchenko, O.N. Grigorashchenko, A.N. Ogurtsov, V.V. Rudenkov, M. Lorenz, M. Frankowski, A.M. Smith-Gicklhorn, V.E. Bondybey, *J. Luminescence,* **94–95,** 475 (2001).
64. E.V. Savchenko, O.N. Grigorashchenko, A.N. Ogurtsov, V.V. Rudenkov, G.B. Gumenchuk, M. Lorenz, M. Frankowski, A.M.Smith-Gicklhorn, V.E.Bondybey, *Surf. Sci*. **507-510,** 754 (2002).
65. E.V. Savchenko, G.B.Gumenchuk, E.M. Yurtaeva, A.G. Belov, I.V. Khizhniy, M. Frankowski, M.K. Beyer, A.M. Smith-Gicklhorn, A.N. Ponomaryov, V.E. Bondybey, *J. Luminescence*, **112,** 101 (2005).
66. A. Ponomaryov, G. Gumenchuk, E. Savchenko and V.E. Bondybey, *Phys. Chem. Chem. Phys*. **9**, 1329 (2007).
67. R.M.J. Cotterill and M. Doyama, *Phys. Letters A* **25**, 35 (1967).
68. B.M. Smirnov, A.S. Yatsenko, *Phys. Usp*., **39,** 211 (1996).
69. C. van de Steen, M. Benhenni, R. Kalus, R. Ćosić, F.X. Gadéa, M. Yousfi. *Plasma Sources Science and Technology*, **28**, 095008 (2019).
70. V.D. Natsik, S.N. Smirnov, Y.I. Nazarenko, *Fiz. Nizk. Temp*. **27,** 1295 (2001)/*Low Temp. Phys*. **27,** 958 (2001); https://doi.org/10.1063/1.1421463.